# Unified Analysis of Cooperative Spectrum Sensing over Composite and Generalized Fading Channels

Ahmed Al Hammadi, Omar Alhussein, *Student Members, IEEE,* Sami Muhaidat, *Senior Member, IEEE,* Mahmoud Al-Qutayri, *Senior Member, IEEE,* Saleh Al-Araji, *Senior Member, IEEE,* George K. Karagiannidis, *Fellow, IEEE*

## Abstract

In this paper, we investigate the performance of cooperative spectrum sensing (CSS) with multiple antenna nodes over composite and generalized fading channels. We model the probability density function (PDF) of the signal-to-noise ratio (SNR) using the mixture gamma (MG) distribution. We then derive a generalized closed-form expression for the probability of energy detection, which can be used efficiently for generalized multipath as well as composite (multipath and shadowing) fading channels. The composite effect of fading and shadowing scenarios in CSS is mitigated by applying an optimal fusion rule that minimizes the total error rate (TER), where the optimal number of nodes is Ahmed Al Hammadi, Mahmoud Al-Qutayri, and Saleh Al-Araji, are with Khalifa University, Abu Dhabi, UAE (e-mails:{100037703, mqutayri, alarajis}@kustar.ac.ae).

Omar Alhussein is with the School of Engineering Science, Simon Fraser University, Burnaby, BC, V5A1S6, (e-mail: oalhusse@sfu.ca).

Sami Muhaidat is with Khalifa University, Abu Dhabi, UAE and with with the School of Engineering Science, Simon Fraser University, Burnaby, BC, V5A1S6, (e-mail: muhaidat@ieee.org).

George K. Karagiannidis is with Khalifa University, Abu Dhabi, UAE, and with Aristotle University, Thessaloniki, Greece (e-mail: geokarag@auth.gr).

derived under the Bayesian criterion, assuming erroneous feedback channels. For imperfect feedback channels, we demonstrate the existence of a TER floor as the number of antennas of the CR nodes increases. Accordingly, we derive the optimal rule for the number of antennas that minimizes the TER. Numerical and Monte-Carlo simulations are presented to corroborate the analytical results and to provide illustrative performance comparisons between different composite fading channels.

**Index Terms**

Energy detection, multipath/composite generalized fading channels, cooperative spectrum sensing, square law selection.

## I. INTRODUCTION

The need for efficient utilization of spectrum has become a fundamental requirement in modern wireless networks, which is mainly due to spectrum scarcity and the ever-increasing demand for higher data rate applications and Internet services [1]. Cognitive radio (CR) networks is a particularly interesting proposal that has been developed to mitigate the spectrum scarcity issue. CR networks can adapt their transmission parameters according to the environment [2]. Cognitive radios have been shown to be very efficient in maximizing spectrum utilization due to their inherent spectrum sensing capability. In a CR network environment, users are categorized as either primary users (PUs) or secondary users (SUs). PUs are the ones who have been assigned spectrum slots, and hence, have higher priority. On the other hand, SUs are opportunistically accessing the under-utilized spectrum. Three detection techniques are commonly used for spectrum sensing in CR networks, namely energy detection (ED), matched filters, and cyclostationary detection [3]. In energy detection, which is the most common detection techniques, due its low implementation complexity, the presence of a PU signal is simply detected by comparing the output of the energy detector with a certain threshold that depends on the SNR [4]. In matched filters, exact information about the transmitted signal waveform, such as its bandwidth and modulation type, is required. Cyclostationary detection, on the other hand, uses statistical properties of the transmitted signals to enhance the probability of PU detection.

Recently, the ED performance over small scale fading channels, such as Rayleigh, Rician, and Nakagami-$m$ is studied in [5], [6]. In addition, unified ED performance analysis over generalized multipath fading channels has been studied by the utilization of the $\kappa - \mu$ and $\kappa - \mu$ extreme fading channels, which outperforms the characterization capabilities of Rician, Nakagami-$m$ and Rayleigh distributions [7]. However, in addition to small scale fading, in most scenarios, the received signal is also degraded by large scale fading (shadowing effect). Therefore, there is a need to analyze the CR performance over composite fading channels, which take into account large scale and small scale fading. It is worth noting that a direct analysis of the ED performance in composite fading channels is rather tedious, since composite fading models can only be represented by infinite integrals. The probability of detection of the ED is derived over Nakagami-lognormal (NL) fading channels in [13]. However, the offered solution is not represented in closed form and the impact of fading and shadowing effects is investigated numerically. Several alternatives that characterize composite fading channels are presented to provide simplified performance analysis for the CR networks. For example, in [8]-[12], the $\mathcal{K}$ distribution is utilized to study the ED performance over RL channels.

One the most effective techniques to mitigate the effect of fading is cooperative spectrum sensing (CSS), where $N$ users sense the spectrum independently and send their soft or hard decisions through an imperfect feedback channel to a fusion center (FC) for a global decision [14]. In CSS, there are two main decision combining schemes, namely soft decision combining, where the SUs send their local observations to the FC, and hard decision combining, where the SUs send their local decisions to the FC. This paper considers the hard decision combining scheme. In [15], CSS optimization over the Rayleigh fading channel has been studied, assuming a perfect feedback channel. In [16], Quan $et.~al.$ studied the performance of CSS under the Bayesian criterion, which accounts for the costs of probabilities of miss-detection and false alarm.

However, the feedback channel is assumed to be perfect and the study is limited to Rayleigh and Suzuki fading channels. In [17], Lee studied the performance of CSS over Rayleigh fading assuming imperfect feedback channels and hard decision combining. Recently, CSS over $\kappa - \mu$ fading channels has been studied in [18]. However, this model does not represent composite fading channels, and the authors considered only the OR fusion rule with a perfect feedback channel.

Although CSS has been investigated extensively in the literature, none of the reported results has considered CSS optimization over composite and generalized fading channels. In this paper, we propose a mixture gamma (MG) based approach to derive a new closed form expression of probability of detection over generalized fading channels. The MG model [19], [31] has been proposed as an alternative to various non-composite and composite fading channels: Lognormal, Weibull, Rayleigh/Lognormal (RL), NL, $\mathcal{K}$, $\mathcal{K}_G$, $\eta-\mu$, $\kappa-\mu$, Hoyt, and Rician channels. The MG model is both accurate and flexible to represent all aforementioned fading channels. It is worth noting that the probability of detection of the MG model has been derived in [19]. However, the solution provided therein is limited to an integer $\beta_k$, which is a parameter in the MG model (4). In this paper, we provide a different approach and derive the probability of detection for both integer $\beta_k$ and arbitrary real $\beta_k$. In addition, the derived probability of detection is extended to show the effect of diversity reception by employing the square-law selection technique (SLS). Moreover, we derive an optimal fusion rule where the optimal number of nodes is derived analytically under Bayesian criterion. Furthermore, we found a TER floor, whereby increasing the number of antennas after this floor will not reduce the TER. As such, we derive the optimal number of antennas that achieves the optimal performance. We stress out that results of this paper are applicable to all fading channels. Our derivations incorporates both cooperative sensing, diversity reception, and assumes erroneous feedback channels. More precisely, our main contributions are summarized as follows:

- The probability of detection of the energy detector using the MG model is derived with and

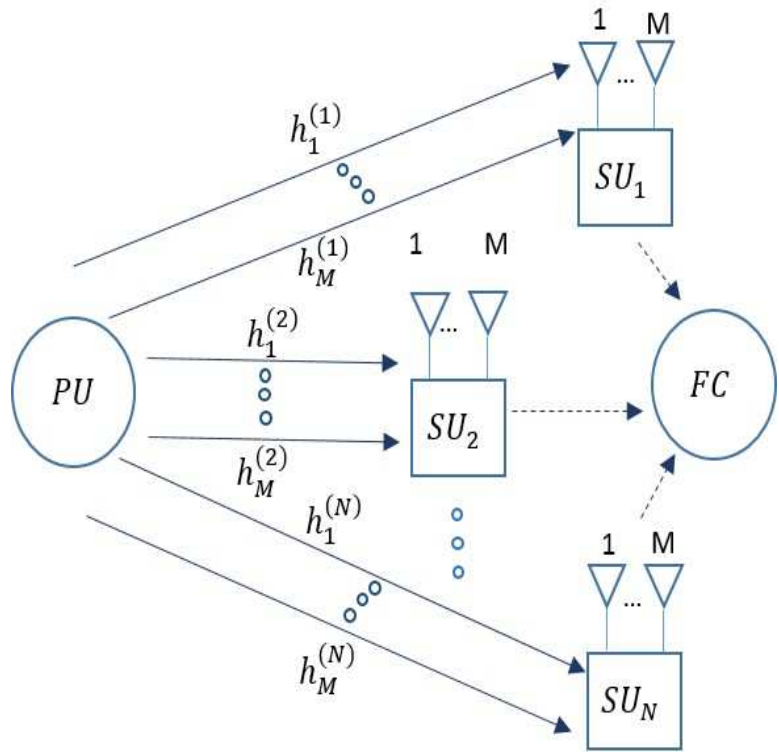

Fig. 1: Cooperative spectrum sensing network configuration.

without diversity reception.

- An optimal fusion rule is derived under Bayesian criterion for CSS optimization over generalized fading channels.
- A closed form expression of the optimal number of antennas $M^*$ is derived.

The rest of the paper is organized as follows. Section II gives a detailed description of the system model, while in Section III, we derive the probability of detection using the MG model. In Section IV, we extend our derivation to diversity reception using the SLS technique. Optimal fusion rule and optimal number of antennas over generalized fading channels are presented in Section V, followed by conclusions in Section VI.

## II. SYSTEM MODEL

We consider a cooperative spectrum sensing network configuration, shown in Fig. 1. The configuration consists of one primary user (PU) and $N$ secondary user (SU) nodes, each equipped with $M$ antennas. The SU nodes report back their hard decisions via imperfect feedback channels to a fusion center (FC), which may represent a base station or a cell site. We assume that the channel gains $h_j^{(i)}$ are independent and identically distributed ($i.i.d$) and are modeled using the generalized MG fading model, where $i = 1, .., N$, and $j = 1, ..., M$.

The received signal copies at the $i^{th}$ SU node and $j^{th}$ antenna can have two possible hypotheses, modeled as

$$\begin{aligned} H_0 &: y_j^{(i)}(t) = v_j^{(i)}(t) \\ H_0 &: y_j^{(i)}(t) = h_j^{(i)}(t)s(t) + v_j^{(i)}(t), \end{aligned} \quad (1)$$

where $H_0$ and $H_1$ represent the absence and presence of the PU, respectively. $s(t)$ corresponds to the transmitted signal from the PU, with energy $\mathbb{E}[|s(t)|^2] = E_s$. $v_j^{(i)}(t) \sim \mathcal{CN}(0, \sigma_n^2)$ is circularly symmetrical complex additive white gaussian noise (AWGN).

Each $i^{th}$ SU node uses an energy detector for the amplitudes $|y_j^{(i)}|_{j=1}^M$ of the received signal and compares it to a threshold $\lambda$. Therefore, the output of this process for each antenna can be written as

$$Z_j^{(i)} = \left|y_j^{(i)}\right|^2 \underset{H_0}{\overset{H_1}{\gtrless}} \lambda, \quad (2)$$

where we have dropped the time index $t$ for simplicity.

For this set-up, the CSS scheme performs following steps:

- Each $i^{th}$ SU calculates the decision statistic for all $M$ antennas and uses the SLS technique to decide on the presence or absence of a PU.
- The binary hard decision of each SU is sent to the FC via an erroneous feedback channel.
- The FC applies a fusion rule to the binary decisions received from all SUs and a final decision is made.

The conditional probabilities of detection and false alarm are calculated using [4]

$$P_d = Q_u(\sqrt{2\gamma}, \sqrt{\lambda_n})$$
$$P_f = \frac{\Gamma(u, \frac{\lambda_n}{2})}{\Gamma(u)}, \quad (3)$$

where $u$ is time-bandwidth product, $Q_u(\cdot,\cdot)$ is the generalized Marcum-Q function [24], $\Gamma(\cdot,\cdot)$ is the upper incomplete gamma function [25, eq. 8.35], $\Gamma(\cdot)$ is the standard gamma function [25, eq. 8.31], $\lambda_n = \lambda/\sigma_n^2$ is the normalized threshold, and $\gamma_j^{(i)} = \frac{\left|h_j^{(i)}\right|^2 E_s}{2\sigma_n^2}$ is the instantaneous SNR of the $i^{th}$ PU-SU link.

Since the probability of false alarm is based on the null hypothesis, it remains the same regardless of the fading scenario. Thus, in the subsequent sections, we focus on the derivation of the probability of detection for the MG distribution with and without diversity.

### III. Probability of Detection over Composite Fading Channels with No Diversity Reception

The MG distribution has been shown to accurately represent several composite and non-composite fading models, and it is written as [19]

$$f_\gamma(x) = \sum_{k=1}^{C} \frac{\alpha_k}{\gamma_0} \left(\frac{x}{\gamma_0}\right)^{\beta_k - 1} \exp\left(-\frac{\zeta_k x}{\gamma_0}\right), \quad (4)$$

where $C$ denotes the number of mixture components. $\alpha_k$ is the mixing coefficient of the $k^{th}$ component having the constraints $\alpha_k > 0$, with $\sum_k \alpha_k = 1$. $\gamma_0$ is the avarage SNR. $\beta_k$ and $\zeta_k$ correspond to the scale and shape parameters of the $k^{th}$ component, respectively. Using the definition, the probability of detection over MG model is given by

$$P_{d,MG} = \sum_{k=1}^{C} \frac{\alpha_k}{\gamma_0} \int_0^\infty Q_u(\sqrt{2x}, \sqrt{\lambda}) \cdot \left(\frac{x}{\gamma_0}\right)^{\beta_k - 1} e^{-\frac{\zeta_k x}{\gamma_0}} dx. \quad (5)$$

Following lemma 1 in [33, eq. 8], and after some simplifications, (5) is solved in exact infinite series representation as follows

$$P_{d,MG} = \sum_{k=1}^{C} \frac{\alpha_k}{\gamma_0} \left[ \sum_{l=0}^{\infty} \frac{(\sqrt{2})^{2l} 2^{\beta_k} \Gamma(\beta_k + 1) \Gamma(u + l, \frac{\lambda}{2})}{l! \Gamma(u + l)(1 + \frac{\zeta_k}{\gamma_0})^{\beta_k + l}} \right]. \quad (6)$$

Although, the above solution is expressed as an infinite series, it converges after relatively few terms. However, for the sake of completeness, it is essential to determine a corresponding simple and accurate truncation error of (6). Following lemma 2 in [33], a closed form upper bound for the truncation error, $\varepsilon_t$, can be derived as

$$\varepsilon_t \leq \sum_{k=1}^{C} \frac{2\alpha_k}{\gamma_0} \Big[ \frac{\Gamma(\beta_k)}{\left(\frac{\zeta_k}{\gamma_0}\right)^{\beta_k}} - \sum_{l=0}^{n} \frac{(\sqrt{2})^{2l} 2^{\beta_k} \Gamma(\beta_k + l)\Gamma(u+l,\frac{\lambda}{2})}{l!\Gamma(u+l)(2 + 2\frac{\zeta_k}{\gamma_0})^{k+l}} \Big]. \tag{7}$$

Additionally, an exact finite series representation is derived given that $\beta_k$ is limited to integer values. With the aid of theorem 1 in [33, eq.3], and after some simplifications, (5) is derived as

$$P_{d,MG}^{int} = \sum_{k=1}^{C} \frac{a_k}{\gamma_0} \Big[ \frac{\Gamma(\beta_k)\Gamma(u,\frac{\lambda}{2})}{\left(\frac{\zeta_k}{\gamma_0}\right)^{\beta_k}\Gamma(u)} + \sum_{l=0}^{\beta_k - 1} \frac{(\lambda)^{4u}\Gamma(\beta_k) {}_1F_1(l+1, u+1, \frac{2\lambda}{4+(4\frac{\zeta_k}{\gamma_0})})}{u!\left(\frac{\zeta_k}{\gamma_0}\right)^{\beta_k-1} 2^u (1+\frac{\zeta_k}{\gamma_0})^{l+1} e^{\frac{\lambda}{2}}} \Big], \tag{8}$$

where the function ${}_1F_1(.,.,.)$ is the confluent hypergeometric function [23, eq. 9.210.1].

In Fig. 2, the complementary receiver operating characteristic (ROC) curves over different non-composite and composite fading channel is plotted using (6). The term 'Numerical Integration' refers to numerically integrating (5) via the trapezoidal integration routine in MATLAB. It can be observed that there is a perfect match between the analytical form derived in (6) and its corresponding simulation, where as shown, the curves span a wide average SNR range. The addressed scenarios were all approximated by two mixture components, $C = 2$. The approximated fading channels and their corresponding MG coefficients can be found in Table I [30].

## IV. Probability of Detection over Composite Fading channels with Diversity Reception

In the SLS scheme, for each $i^{th}$ SU node, the branch with the maximum $\gamma_j$ is selected as follows [5]

$$\gamma_{SLS} = \max_j(\gamma_j), j = 1,..,M, \tag{9}$$

TABLE I: MG coefficients for different composite and non composite fading channels

| Channel | Param | $\alpha$ | $\beta$ | $1/\zeta$ | MSE |
|---|---|---|---|---|---|
| Lognormal | $\zeta = 1$ | 0.8795306 | 21.47391625 | 0.04634621 | 2.18E-04 |
|  |  | 0.1204694 | 19.29669267 | 0.06526572 |  |
| Weibull | $m = 4$ | 0.4163066 | 2.547627 | 0.276866 | 8.14E-05 |
|  |  | 0.5836934 | 6.410809 | 0.1887993 |  |
| Rayleigh | - | 0.2803494 | 0.9124631 | 0.4795083 | - |
|  |  | 0.7196506 | 1.3046339 | 0.9339172 |  |
| Nakagami-m | - | $\frac{m^m}{\Gamma(m)}$ | $m$ | $\frac{1}{m}$ | - |
| Ray/logn | $\zeta = 1$ | 0.2889985 | 0.9667361 | 3.908869 | 1.18E-06 |
|  |  | 0.7110015 | 0.972492 | 1.62796 |  |
| Ray/logn | $\zeta = 1.5$ | 0.7229848 | 0.9223047 | 0.8351385 | - |
|  |  | 0.2770152 | 0.8499458 | 3.0324419 |  |
| Ray/logn | $\zeta = 0.5$ | 0.3491298 | 0.9117919 | 0.6304104 | 1.43E-06 |
|  |  | 0.6508702 | 1.225578 | 1.010205 |  |
| Nak/logn | $m = 2$ | 0.6569638 | 1.9707105 | 0.4273802 | 1.55E-06 |
|  | $\zeta = 0.5$ | 0.3430362 | 2.5034565 | 0.5267421 |  |
| Nak/logn | $m = 4$ | 0.7775037 | 4.0356456 | 0.2247047 | 3.76E-06 |
|  | $\zeta = 0.5$ | 0.2224963 | 5.2938705 | 0.2556295 |  |

where $M$ denotes the number of antennas at each SU node. Under $H_0$, the probability of false alarm for the SLS scheme can be expressed as

$$P_{f,SLS} = 1 - \Pr(\gamma_{SLS} < \lambda_n | H_0). \qquad (10)$$

Substituting (9) in (10), we obtain

$$P_{f,SLS} = 1 - \Pr(\max(\gamma_1, \gamma_2, .., \gamma_M) < \lambda_n | H_0). \qquad (11)$$

Using order statistics [29], $P_{f,SLS}$ is written as

$$P_{f,SLS} = 1 - [1 - P_f]^M. \qquad (12)$$

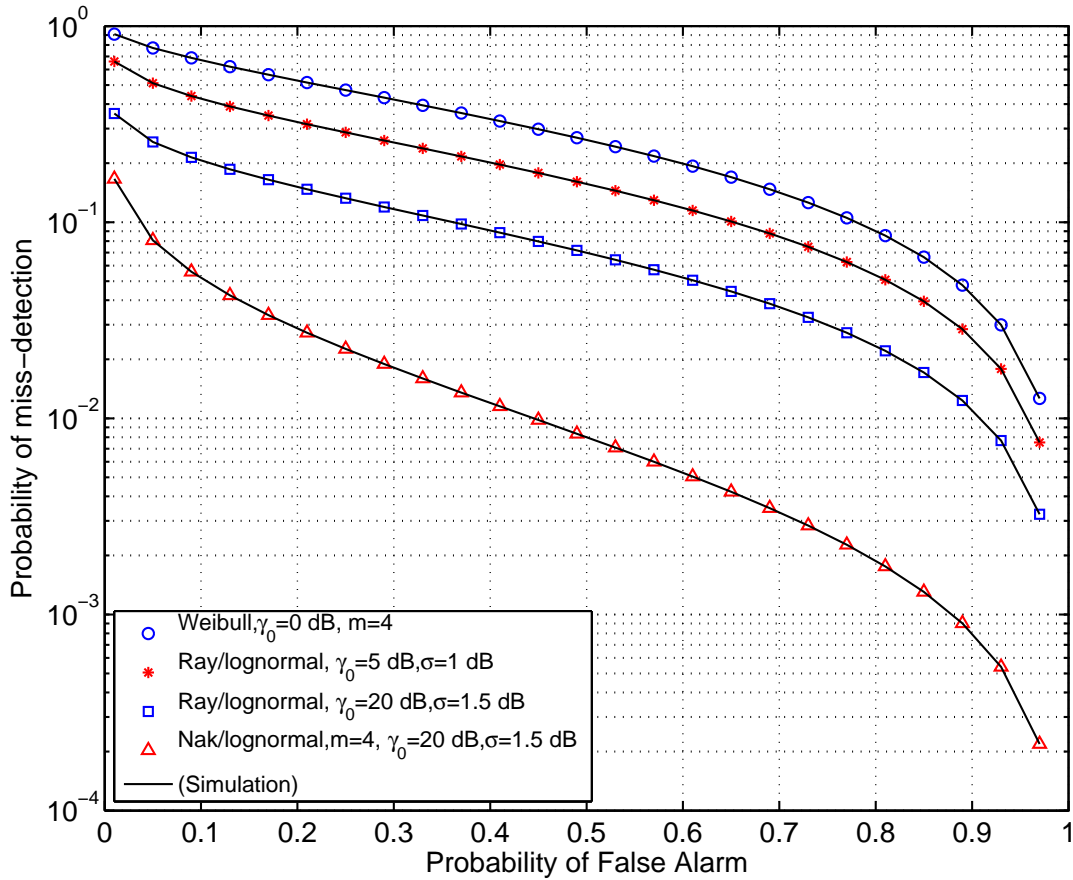

Fig. 2: Complimentary ROC for single SU over various fading channels.

Similarly, the unconditional probability of detection over AWGN channel is obtained by

$$P_{d,SLS} = 1 - \prod_{j=1}^{M} \left[ 1 - Q_u(\sqrt{2\gamma_j}, \sqrt{\lambda_n}) \right]. \tag{13}$$

Averaging (13) over (6), we obtain the unconditional probability of detection under the SLS scheme, $\bar{P}_{d,SLS}$, expressed as

$$\bar{P}_{d,SLS} = 1 - \prod_{j=1}^{M} [1 - P_{d,MG}]. \tag{14}$$

## V. Optimal fusion rule and optimal number of antennas in Composite Fading Channels

Cooperative spectrum sensing can highly decrease the probability of miss-detection in fading channels. In this section, the performance of cooperative spectrum sensing is studied over various fading channels. Practically, the channel between the SUs and the FC is imperfect. Therefore, the hard decisions sent by each SU node are affected by the quality of the feedback channel. Here, a binary symmetric channel with error probability $q$ is assumed. The probabilities of detection and false alarm over erroneous feedback channels are given by [17]

$$P'_f = (1-q)P_f + q(1-P_f), \quad (15)$$

$$P'_{d,MG} = (1-q)(P_{d,MG}) + (q)(1-P_{d,MG}). \quad (16)$$

For the SLS diversity reception, we substitute (12) and (14) in (15) and (16) respectively, yielding

$$P'_{f,SLS} = (1 - (1-P_f)^M)(1-q) + (1-P_f)^M q, \quad (17)$$

and

$$P'_{d,SLS} = (1-q)\left(1 - \prod_{j=1}^{M}(1-P_{d,MG})\right) + q\prod_{j=1}^{M}(1-P_{d,MG}). \quad (18)$$

Assuming $i.i.d$ diversity branches, (18) can be expressed as

$$P'_{d,SLS} = (1-q)\left(1 - (1-P_{d,MG})^M\right) + q(1-P_{d,MG})^M. \quad (19)$$

The probability of false alarm $Q_f$ and the probability of detection $Q_d$ at the FC using $k$-out-of-$N$ rule is thus given by

$$Q_f = \sum_{n=k}^{N} \binom{N}{n} (P'_{f,SLS})^n (1 - (P'_{f,SLS}))^{N-n}$$
$$= 1 - B_F(k-1, N, P'_{f,SLS}), \quad (20)$$

and

$$Q_d = \sum_{n=k}^{N} \binom{N}{n} (P'_{d,SLS})^n (1 - (P'_{d,SLS}))^{N-n}$$

$$= 1 - \mathrm{B}_F(k-1, N, P'_{d,SLS}), \qquad (21)$$

where $B_F$ is the binomial cumulative distribution function [26]. Since $P_m = 1 - P_d$, the probobility of miss-detection at the FC is given by

$$Q_m = \sum_{n=k}^{N} \binom{N}{n} (1 - P'_{d,SLS})^n (P'_{d,SLS})^{N-n}$$

$$= \mathrm{B}_F(k-1, N, 1 - P'_{d,SLS}). \qquad (22)$$

### A. Optimal Fusion Rule under Bayesian Criterion

The performance of CSS in composite and non-composite fading environments can be improved by applying an optimal fusion rule to decrease the TER. The problem is formulated under the Bayesian criterion which accounts for the cost of miss-detection and false alarm. Such cost can be set by a regulatory or based on past experiences which makes the preceding analysis both practical and widely applicable to any fading channel. We consider the problem of minimizing the Bayesian risk of CSS, which is given by [16]

$$\min_{1 \leq k \leq N} R(k) = W_f Q_f + W_m Q_m, \qquad (23)$$

where $W_m$ and $W_f$ are the cost of miss-detection and false alarm, respectively. From (20) and (22), the Bayesian risk $R(k)$ can be written as

$$R(k) = W_f - W_f B_F(k-1, N, P'_{f,SLS})$$

$$+ W_m B_F(k-1, N, P'_{d,SLS}), \qquad (24)$$

Let $R_0(k)$ be defined as

$$R_0(k) \triangleq W_f B_F(k-1, N, P_f) - W_m B_F(k-1, N, P_d), \qquad (25)$$

then we have

$$\frac{\partial R(k)}{\partial k} = \binom{N}{k} [W_m (P'_{d,SLS})^k (1-P'_{d,SLS})^{N-k}$$
$$- W_f (P'_{f,SLS})^k (1-P'_{f,SLS})^{N-k}]. \quad (26)$$

Thus, the optimal fusion rule, $k_{opt}$, which minimizes the Bayesian risk can be obtained as

$$k_{opt} \approx \left\lceil \frac{\ln\left(\frac{W_m}{W_f} \cdot \frac{1-P'_{f,SLS}}{1-P'_{d,SLS}}\right)^N}{\ln\left(\frac{(P'_{d,SLS})(1-P'_{f,SLS})}{(1-P'_{d,SLS})(P'_{f,SLS})}\right)} \right\rceil, \quad (27)$$

where $\lceil . \rceil$ denotes the ceiling function.

Fig. 3 shows the TER as a function of the normalized threshold $\lambda_n$ over NL and RL composite fading channels using different fusion rules. It is clear from the graph that the optimal fusion rule yields the best performance for both NL and RL channels. The minimum TER of the optimal fusion rule ($k = 3$) in the NL channel is $10^{-3.9}$, whereas in the RL channel it is lower by an order of magnitude which is $10^{-2.9}$. On the other hand, the AND rule yields the worst performance, yielding a minimum TER of $10^{-1}$ for the RL scenario and $10^{-2}$ for the NL scenario. Interestingly, the OR rule did not improve from RL to NL and the minimum TER is $10^{-2}$ for both channels. Therefore, we conclude that using optimal fusion rule always increases the performance as compared to other fusion rules.

In Fig. 4, we employ the same fading scenario with having different weights for miss-detection and false alarm, $M = 2$ and $\gamma_0 = 5$ dB. As observed, the optimal fusion rule holds its optimality with a minimum TER of $10^{-4.5}$ for the NL channel and $10^{-3.8}$ for the RL channel. The OR rule improved slightly with a minimum TER of $10^{-2.8}$. Finally, the performance of AND rule is still the worst amongst all the fusion rules in both channels.

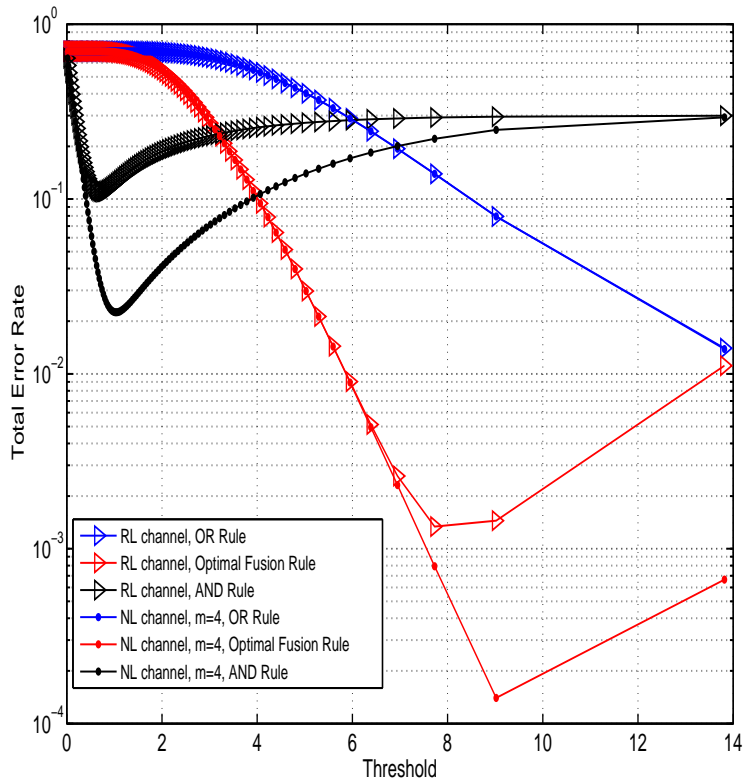

Fig. 3: Total error rate vs normalized detection threshold over the NL channel using different fusion rules with $\gamma_0 = 10$ dB , $m = 4$, $\zeta = 0.5$ dB, $M = 1$, $N = 10$, $W_m = 0.3$, $W_f = 0.7$ , and $q = 10^{-2}$.

## B. Optimal Number of Antennas

In this section, we investigate the effect of increasing the number of antennas with SLS diversity reception. It has been noted that SLS can increase the performance of the CSS remarkably. Here, we found out that the performance is not monotonically increasing if the feedback channel is erroneous. Fig. 5 shows the effect of the feedback channel on the TER while increasing number of antennas $M$ in various composite fading channels. In the perfect feedback channel case, the TER is monotonically decreasing as we increase the number of antennas. However, in the imperfect feedback channel case, the total error rate decrease until a specific lower bound whereby increasing $M$ afterwards is ineffective and inefficient. Since the function in the imperfect feedback channel case is convex, the optimal number of antennas $M^*$ can be evaluated when

$$\frac{\partial Q_m}{\partial M} + \frac{\partial Q_f}{\partial M} = 0 \qquad (28)$$

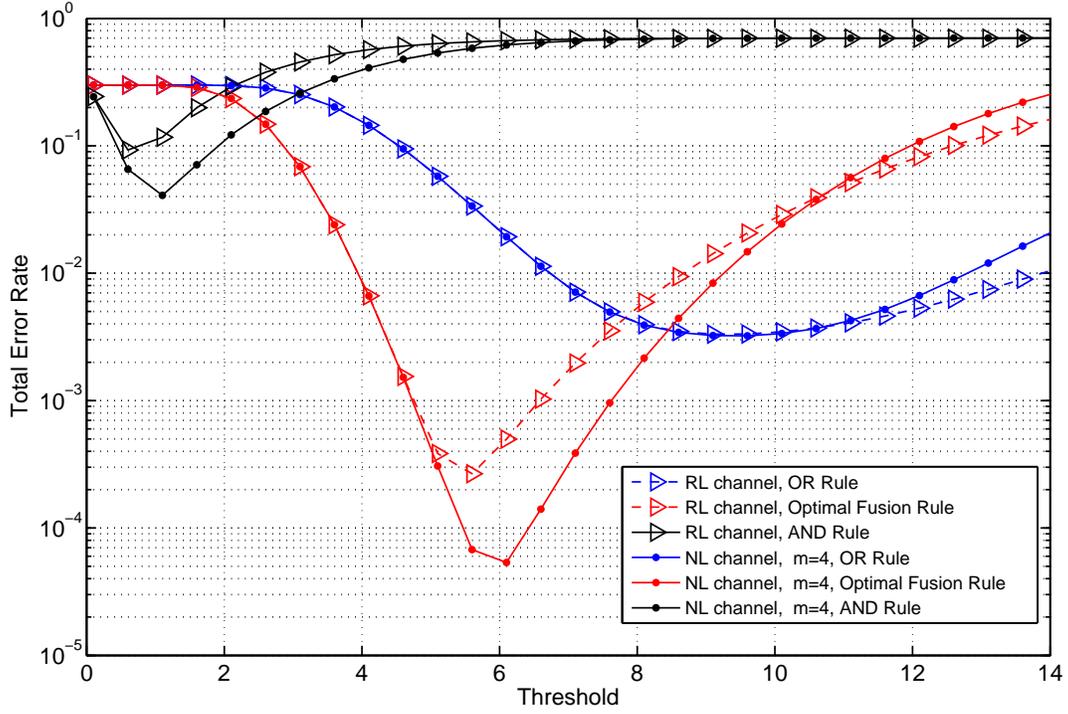

Fig. 4: Total error rate vs normalized detection threshold over the RL channel using different fusion rules with $\gamma_0 = 5$ dB, $\zeta = 0.5$ dB, $M = 2$, $N = 10$, $W_m = 0.7$, $W_f = 0.3$ , and $q = 10^{-2}$.

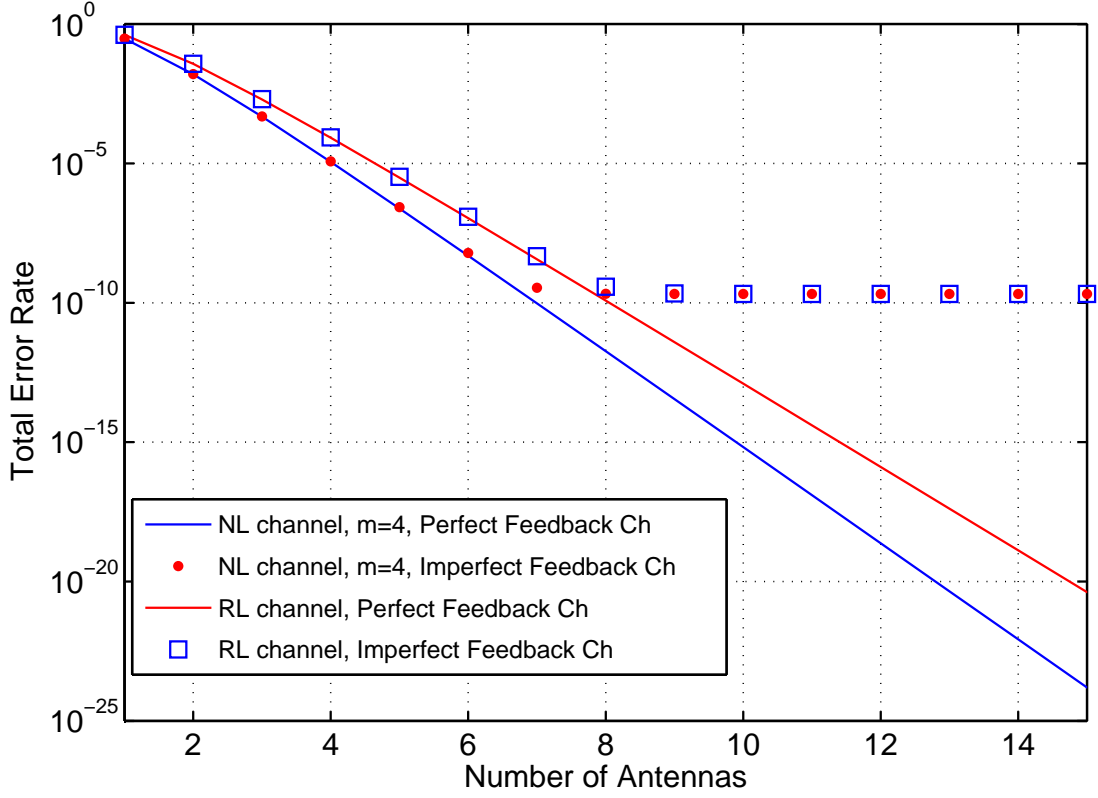

Fig. 5: Total error rate vs number of antennas over perfect and imperfect feedback channels with $\gamma_0 = 5$ dB, $\zeta = 0.5$ dB, $k = 3$, $N = 10$, $\lambda_n = 8$, $W_m = 0.3$, $W_f = 0.7$, and $q = 10^{-2}$.

where

$$\frac{\partial Q_f}{\partial M} =$$

$$\sum_{n=k}^{N} \binom{N}{n} n (P'_{f,SLS})^{n-1} \frac{\partial P'_{f,SLS}}{\partial M} (1 - P'_{f,SLS})^{N-n}$$

$$- \sum_{n=k}^{N} \binom{N}{n} (P'_{f,SLS})^{n} (N-n)(1 - P'_{f,SLS})^{N-n-1}$$

$$\times \frac{\partial P'_{f,SLS}}{\partial M}$$

$$= \frac{\partial P'_{f,SLS}}{\partial M} \cdot \sum_{n=k}^{N} \binom{N}{n} (P'_{f,SLS})^{n-1} (1 - P'_{f,SLS})^{N-n}$$

$$\times \left[ (n - (N-n) \cdot \frac{P'_{f,SLS}}{1 - P'_{f,SLS}} \right], \quad (29)$$

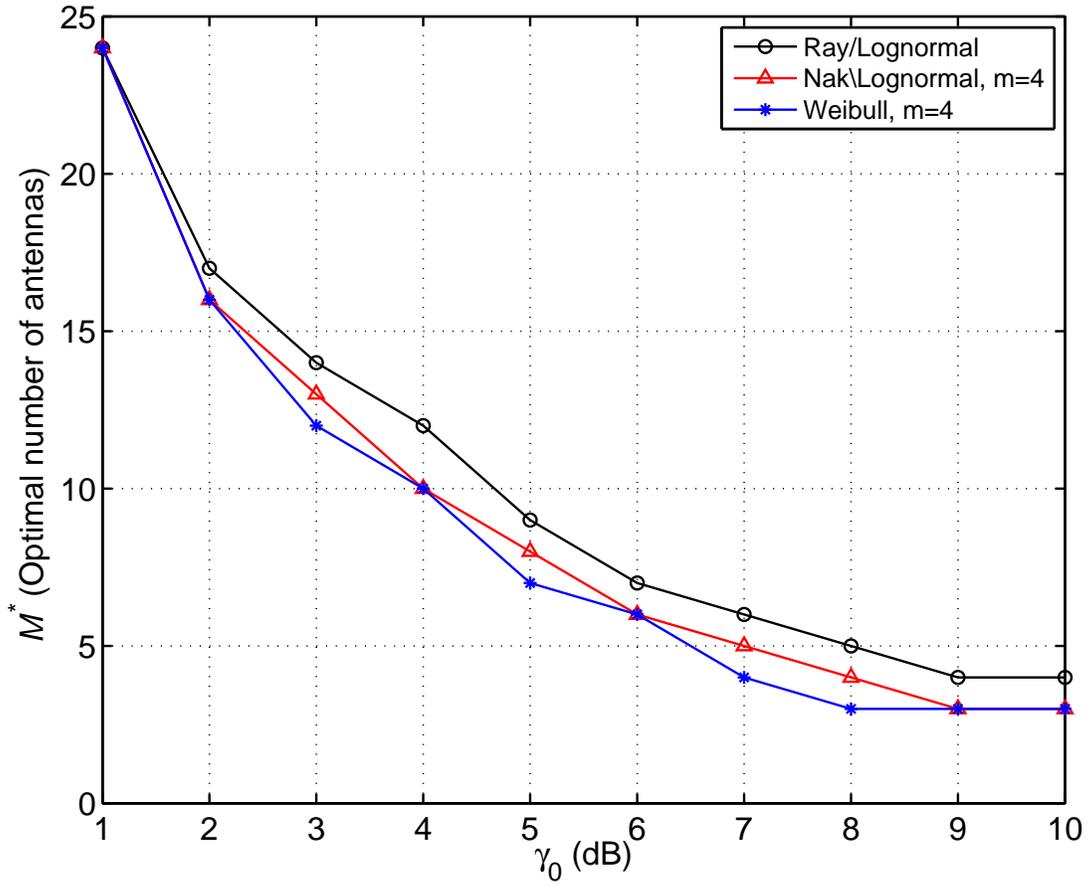

Fig. 6: Optimal number of antennas vs. average SNR in various fading channels with, $k = 3$, $N = 10$, $\lambda_n = 8$, $W_m = 0.7$, $W_f = 0.3$ , and $q = 10^{-2}$.

and

$$\frac{\partial P'_{f,SLS}}{\partial M} = (1 - P_f)^M (-1 + 2q) \ln(1 - P_f). \tag{30}$$

Similarly

$$\frac{\partial Q_m}{\partial M} =$$

$$-\sum_{n=k}^{N} \binom{N}{n} n(\bar{P}'_{d,SLS})^{n-1} \frac{\partial \bar{P}'_{d,SLS}}{\partial M}(1 - P'_{d,SLS})^{N-n}$$

$$+\sum_{n=k}^{N} \binom{N}{n} (\bar{P}'_{d,SLS})^n (N-n)(1-\bar{P}'_{d,SLS})^{N-n-1}$$

$$\times \frac{\partial \bar{P}'_{d,SLS}}{\partial M}$$

$$= -\frac{\partial \bar{P}'_{d,SLS}}{\partial M} \sum_{n=k}^{N} \binom{N}{n} (\bar{P}'_{d,SLS})^{n-1}(1-\bar{P}'_{d,SLS})^{N-n}$$

$$\times \left[(n - (N-n)\cdot\frac{\bar{P}'_{d,SLS}}{1-\bar{P}'_{d,SLS}}\right], \tag{31}$$

and

$$\frac{\partial P'_{d,SLS}}{\partial M} = (1 - P_{d,MG})^M q \ln(1 - P_{d,MG})$$

$$- (1 - P_{d,MG})^M (1-q) \ln(1 - P_f)$$

$$= (1 - P_{d,MG})^M (-1 + 2q) \ln(1 - P_{d,MG}). \tag{32}$$

The optimal number of antennas can be obtained by substituting (29) and (31) in (28) and solving it for $M$ numerically.

Fig. 6 shows the optimal number of antennas as function of the avarage SNR, over various composite and non-composite fading channels. As expected, the optimal number of antennas is inversely proportional to $\gamma_0$. The RL channel requires in general more antennas than the NL and Weibull channels to reach $M^*$, since the NL and Weibull channels depicts mild multipath fading with $m = 4$.

## VI. Conclusions

We have investigated the performance of cooperative spectrum sensing over generalized fading channels. A unified MG based approach is proposed for the performance analysis of cooperative spectrum sensing in cognitive radio. It has been found that in a composite fading channel, the performance can be improved remarkably by applying an optimal fusion rule outperforming the AND rule. Moreover, increasing the number of antennas in a CSS network with erroneous feedback channel does not decrease the TER monotonically. Based on this observation, the optimal rule for number of antennas is determined numerically. Finally, it has been found that the fading severity can highly affect the optimal rule for the number of antennas.